\documentstyle[12pt]{article}

\newcommand{\mpl}{M_{\rm pl}}
\newcommand{\mst}{M_{*}}

\newcommand{\be}{\begin{equation}}
\newcommand{\ee}{\end{equation}}
\newcommand{\bea}{\begin{eqnarray}}
\newcommand{\eea}{\end{eqnarray}}

\newcommand{\newc}{\newcommand}
\newc{\gsim}{\lower.7ex\hbox{$\;\stackrel{\textstyle>}{\sim}\;$}}
\newc{\lsim}{\lower.7ex\hbox{$\;\stackrel{\textstyle<}{\sim}\;$}}
\newc{\gev}{\,{\rm GeV}}
\newc{\mev}{\,{\rm MeV}}
\newc{\ev}{\,{\rm eV}}
\newc{\kev}{\,{\rm keV}}
\newc{\tev}{\,{\rm TeV}}

\renewcommand{\phi}{\varphi}
\newc{\smu}{{\tilde\mu}}
\newc{\snu}{{\tilde\nu}}
\newc\order{{\cal O}}
\newc{\eps}{\epsilon}
\newc{\re}{\mbox{Re}\,}
\newc{\im}{\mbox{Im}\,}
\newc{\lunits}{\,\mbox{cm}^{-2}\mbox{s}^{-1}}
\relax
\def\boxeqn#1{\vcenter{\vbox{\hrule\hbox{\vrule\kern3pt\vbox{\kern3pt
\hbox{${\displaystyle #1}$}\kern3pt}\kern3pt\vrule}\hrule}}}
\def\qed#1#2{\vcenter{\hrule \hbox{\vrule height#2in
\kern#1in \vrule} \hrule}}
\newc{\ie}{{\it i.e.}}          \newc{\etal}{{\it et al.}}
\newc{\eg}{{\it e.g.}}          \newc{\etc}{{\it etc.}}
\newc{\cf}{{\it c.f.}}
\def\ltap{\ \raise.3ex\hbox{$<$\kern-.75em\lower1ex\hbox{$\sim$}}\ }
\def\gtap{\ \raise.3ex\hbox{$>$\kern-.75em\lower1ex\hbox{$\sim$}}\ }
\def\gl{\ \raise.5ex\hbox{$>$}\kern-.8em\lower.5ex\hbox{$<$}\ }
\def\roughly#1{\raise.3ex\hbox{$#1$\kern-.75em\lower1ex\hbox{$\sim$}}}
\def\dsl{\,\raise.15ex\hbox{/}\mkern-13.5mu D} 
\def\delsl{\raise.15ex\hbox{/}\kern-.57em\partial}
\def\Ksl{\hbox{/\kern-.6000em\rm K}}
\def\Asl{\hbox{/\kern-.6500em \rm A}}
\def\Dsl{\hbox{/\kern-.6000em\rm D}} 
\def\Qsl{\hbox{/\kern-.6000em\rm Q}}
\def\gradsl{\hbox{/\kern-.6500em$\nabla$}}
\let\beta=\beta

\let\Ga=\Gamma
\let\de=\delta

\newdimen\pmboffset
\def\oldpmb#1{\setbox0=\hbox{#1}%
 \copy0\kern-\wd0
 \kern\pmboffset\raise 1.732\pmboffset\copy0\kern-\wd0
 \kern\pmboffset\box0}

\def\vev#1{\left\langle #1 \right\rangle}

\def\inv{^{\raise.15ex\hbox{${\scriptscriptstyle -}$}\kern-.05em 1}}
\def\pr#1{#1^\prime}  
\def\lbar{{\lower.35ex\hbox{$\mathchar'26$}\mkern-10mu\lambda}} 

\let\<=\langle
\let\>=\rangle

\let\+=\uparrow
\let\-=\downarrow

\begin{document}

\baselineskip=17pt
\pagestyle{plain}
\setcounter{page}{1}

\begin{titlepage}

\begin{flushright}
CERN-TH/99-56\\
SU-ITP-99/12

\end{flushright}
\vspace{10 mm}

\begin{center}
{\LARGE Early Inflation and Cosmology in Theories}
\vskip 2mm
{\LARGE with Sub-Millimeter Dimensions\footnote{Talk
presented by Savas Dimopoulos at COSMO98.  The
work of SD an NK is supported in
part by NSF grant PHY-9870115. The work of JMR is supported
in part by an A.P. Sloan Foundation Fellowship.}}
\vspace{3mm}
\end{center}
\begin{center}
{\large Nima Arkani-Hamed$^a$, Savas Dimopoulos$^b$, Nemanja
Kaloper$^b$,}\\ 
{\large and John March-Russell$^c$}\\ \vspace{3mm} {\em 
$^a$ SLAC, Stanford University, Stanford, CA 94309, USA}\\ 
{\em $^b$
Physics Department, Stanford University, Stanford CA 94305, USA}\\
{\em $^c$ Theory Division, CERN, CH-1211, Geneva 23, Switzerland}
\end{center}
\vspace{3mm}
\begin{center}
{\large Abstract}
\end{center}
\noindent
We discuss early cosmology in theories where the fundamental Planck mass 
is close to the TeV scale.  In such theories the standard model fields are
localized to a $(3+1)$-dimensional wall with 
$n$ new transverse sub-millimeter sized spatial dimensions.  
The topic touched upon include: early inflation that occurs
while the size of the new dimensions are still small,
the spectrum and magnitude of density perturbations, the post-inflation
era of contraction of our world while the internal dimensions
evolve to their final ``large'' radius, and the production of
gravitons in the bulk during these two eras.  The radion moduli
problem is also discussed.
\end{titlepage}
\newpage


Following Refs.\cite{ADD,AADD,ADDlong}, there has been much recent interest
in the possibility that the fundamental Planck scale $\mst$
is close to a $\tev$~\cite{antoniadis,other,AHDMR,new,cosmo,kl,dt}.
The weakness of gravity at long
distances is then explained by $n$ new ``large" dimensions, the (present)
size, $b_0$, of which is given by the Gauss Law relation
$\mpl^2 =(b_0)^n \mst^{n+2}$ where $\mpl=2\times 10^{18}\gev$.  
Taking $\mst \sim 1\tev$ implies the case $n=1$ is excluded since
$b_0 \sim 10^{13}$ cm, while for $n\ge2$ however, $b_0 \ltap 1$
mm, which is the distance where our present experimental
knowledge of gravitational strength forces ends.  
The success of the standard model (SM) up to $\sim 100\gev$ implies
that the SM fields can not feel these extra dimensions; they must be
localized to a 3d wall, or ``3-brane", in the
higher dimensional space. 
The most attractive possibility for localizing the SM fields to
the brane is to employ the D-branes that naturally occur in
string theory~\cite{Dbrane,AADD}; this approach has the
advantage of being formulated within a consistent quantum
theory of gravity.
However, the most important question is whether this framework is
experimentally invalidated: In Ref.~\cite{ADDlong}, laboratory,
astrophysical, and cosmological constraints were found not to
exclude these ideas.

Here we will summarize our understanding of inflation and other
features of early universe cosmology in such
models~\cite{AHDKMR}.\footnote{See also the
companion contribution~Ref.\cite{nemanja}.}
In particular we are concerned with three very important questions:
1) How does inflation arise in these models, or is some
alternative necessary?  2) How do the new spatial dimensions
evolve to their current stabilized value?  3) Does either
inflation or the internal evolution produce too much matter
in the bulk?  Additionally, is there an equivalent of the
Polonyi or moduli problem?

\section*{Inflation}

The most attractive models of inflation
occur while the internal dimensions are still small, far away from their
final stabilized value $b_0$~\cite{AHDKMR}.\footnote{In particular,
we avoid all the problems~\cite{cosmo,kl}
which are met if inflation occurs {\it after} the new dimensions
reach their current, large, size.  Also note that post-stabilization
inflation cannot explain the age of the universe as 
wall-only inflation cannot begin before $t \sim H^{-1} \sim
\mpl/\mst^2 >> \mst^{-1}$, when the universe is already very large
and old~\cite{kl,nemanja}.}  Concretely, the basic story is:

\begin{itemize}
\item
The quantum creation of the universe takes place with the initial
size of all dimensions close to the fundamental Planck scale
$\mst^{-1}$. 
\item
A prolonged period of inflation in a direction parallel to
our brane takes place.  The approximately scale-invariant
nature of the observed primordial perturbation spectrum implies that, 
during inflation, the internal dimensions must expand more slowly than 
the universe on the wall.  Thus we are led to consider
a form of asymmetric inflationary expansion of the
higher-dimensional world ($b(t)=b_I$ is essentially static). 
Since the internal dimensions are small the effective
4-dimensional Newton's constant is large:
$G_{\rm N,initial} = 1/b_I^n \mst^{n+2} \simeq 1/\mst^2$.
\item
Thus the Hubble constant during this initial period of inflation
can be {\it large} even though the energy density is quite small,
$\vev{V}\le \tev^4$,
\be
H_{\rm infl}^2 \simeq {\vev{V}\over 6 b_I^n \mst^{(n+2)}}
\simeq {\vev{V}\over \mst^2}.
\ee
Therefore inflation can be {\it rapid}, and moreover, the density
perturbations can be large, being determined to be
\be
{\de\rho\over \rho} = \frac{5}{12\pi\sqrt{2n(n-1)}}
{ H_{\rm infl}\over \mst (\mst b_I)^{n/2} S },
\ee
where $S$ is a potential-dependent parameter that
encapsulates both the duration
of $a(t)$ inflation and the deviation of the
perturbation spectrum from scale-invariant Harrison-Zeldovich.
(We argue that $S\ltap 1/50$.) 
\item
Specifically, the deviation of the spectral index $n_\rho$ of density
perturbations $\delta\rho/\rho$ from scale invariance is given by:
\be
n_\rho -1 \simeq -{n(n+2)\over 2}\left( S^2 +
ST (b/b_I -1)^2 \right),
\label{nrhofinal}
\ee
where $T$ is another potential-dependent parameter.  Thus
to have sufficient scale invariance we need just mildly
tune $S,T\ltap 1/50$, or alternatively $S\ltap 10^{-3}$ and $T\sim 1$.
Similarly, the number of efolds is given by
\be
N_e \simeq {1\over S+T}\left( {2\sqrt{T}\tan^{-1}(1/\sqrt{S})\over
\sqrt{S}} - \log(1+1/S) + 2 \log(1+1/\sqrt{T}) \right)
\label{efoldcond}
\ee
which for $N_e\gtap 100$ is actually slightly more stringent
than (\ref{nrhofinal}).
\item
By going to an effective 4d theory on the wall where
$b(t)$ appears as a Brans-Dicke like field (in Einstein frame),
it is possible to see that the conditions on $S$ and $T$
are nothing but {\it the usual slow-roll conditions}
for a scalar-gravity theory.
\item
In the minimal approach, the inflaton field is just the
moduli describing the size of the new dimensions
(the radion field of \cite{AHDMR}), the role
of the inflationary potential being played by the
stabilizing potential of this internal space.
In the case of a wall-localized inflaton, this early
inflation might even result from the electroweak phase
transition, in which case the inflaton is the Higgs.
Actually, an important remark in this regard is that when
the internal dimensions are small $b_I \sim \mst^{-1}$ the
distinction between on-the-wall and off-the-wall physics
is not meaningful: \eg, the inflationary features in $V(b)$
at small $b$ could be {\it due} to Higgs physics on the wall. 
\end{itemize}

\section*{Evolution towards stabilization point\footnote{This
section contains material in addition to that presented at COSMO98.}}

Under quite general conditions,
the inflationary era is followed by an
epoch where the scale factor of our brane-universe undergoes
a slow {\it contraction} while the internal dimensions
expand towards their final stabilized value.
Even with the inclusion of a potential for $b$, it
is possible to exactly solve for the evolution during this epoch,
which generalizes the usual vacuum Kasner solutions.  
The history can be summarized as:
\begin{itemize}
\item
Wall inflation ends, and simultaneously
the radion starts to evolve to its' minimum at $b_0$.  
Almost immediately our scale factor $a(t)$ begins to contract.  
Both $a$ and $b$ have determined (subluminal) power-law
dependence on time.
\item
At generic values of $b$ away from the
stabilization point the potential $V(b)$ can be approximated by
$V = W b^{-p}$ ($V$ is the effective 4d potential,
$[W] = 4-p$).\footnote{An additional logarithmic dependence
of $V$ on $b$ is quite possible and harmless.} 
Then the asymptotic form of the exact solutions
places an upper bound on the amount of contraction of the brane 
as a function of $b$: 
\be
\frac{a_f}{a_i} \le \left(\frac{b_i}{b_f}\right)^\zeta
\label{contrrad}
\ee
where the parameter $\zeta$ is given by
($\Delta = 6n - 4np - n^2 - p^2$)
\be
\zeta = \frac{n(n+p-2)}{2(2n+p)} ~~~{\rm for} ~\Delta > 0,\qquad
\zeta = \frac{3n-\sqrt{3n(n+2)}}{6} ~~~ {\rm for} ~\Delta < 0.
\label{contrrad2}
\ee
\item
Remarkably this epoch of contraction {\it ends}. 
There are two generic possibilities for how this
can happen:  The first involves the reheating of the
wall.  Contraction of $a(t)$ stops and reverses
when $\rho_{wall}$ satisfies
\be
\rho_{wall}= n(n-1) \mst^{n+2} b^n H^2_b \equiv \rho_b
\ee
Here $\rho_b$ is the radion (kinetic) energy.  Interestingly
there is a model-independent form of such reheating that results
from the
primordial $\rho_{wall}$ left over from the inflationary
epoch.  If there is sufficient $a(t)$ contraction,
this blue-shifting de~Sitter phase remnant $\rho_{wall}$ can become
comparable to $\rho_b$ {\it before} $b$ reaches the stabilization
point, $b_0$.  It is also possible that a form of reheating takes place
on the wall which is totally unconnected with the contraction of $a(t)$,
but that again leads to $\rho_{wall} \ge \rho_b$.
Possibilities in this class include the decay of some metastable
state on the wall, or the collision of some other brane
with our brane.
After the contraction of $a(t)$ reverses, the 
radion and wall-localized energy densities scale together
until the stabilization point is reached.
The second case is where $b(t)$ reaches the stabilization point
before $\rho_{wall}=\rho_b$.  By going to the Einstein frame it is
possible to show that once stabilization of $b$ occurs, the period
of $a(t)$ contraction automatically ceases.
\item
The total amount of 
contraction of $a(t)$ is bounded 
and varies between at most $7$ efoldings in the case
of $n=2$ to at most $12$ efoldings when $n=6$.
\end{itemize} 

\section*{Bulk graviton production, and moduli problem?}

Naively one would worry that during the era of radion evolution
the bulk becomes full of (dangerous) gravitons.  
There are actually two slightly different issues here:
i) The energy density in Kaluza-Klein (KK) excitations of the
graviton in the bulk, and, ii) the energy density in the (would-be)
{\em zero mode} of the bulk graviton, namely the radion.  
The constraints on the energy density in KK excitations
is actually more severe than that on the radion energy
density, which just comes from overclosure.  The
reason for this is the diffuse gamma-ray background constraint
arising from KK decays~\cite{ADDlong}.  Again an executive
summary of the relevant arguments is:

\begin{itemize}
\item
A time-dependent gravitational
field can produce particles from the vacuum.  
Including this, the equation for the effective 4d
KK energy density becomes
\be
\dot \rho_{KK} + 3H_a \rho_{KK} + H_b \rho_{KK} = H^5 .
\ee
\item
From this, and the expressions for $a(t)$ and $b(t)$ during the contraction
phase, one can show that the {\it dominant contribution to bulk
KK graviton production arises from early times}.   In fact
$\rho_{KK,{\rm final}}$
differs from the final energy density of the blue-shifted
wall-localized radiation at the end of the epoch of contraction only
in that it is further suppressed by a factor of $(a_f/a_i)(b_i/b_f)$, 
which comes from the fact that the KK gravitons are red-shifted
by the {\it bulk} expansion, and only concentrated and not blue-shifted
by the wall contraction.  Thus we get
\be
\frac{\rho_{\rm KK,f}}{\rho_{\rm wall,f}} \le  
\left(\frac{a_f b_i}{a_i b_f}\right)
\ltap \left(10^{1-30/n}\right)^{1 + \zeta},
\label{gravbound}
\ee
using the conservative estimates $b_i \sim 10 \mst^{-1}$ and $b_f=b_0$.
Evaluating this in, for example, the case of the simple Kasner
contraction with exponents given in (\ref{contrrad2}) leads
to $\rho_{KK}/\rho_{wall}$ varying between $3\times 10^{-17}$
for $n=2$ to $1\times10^{-8}$ for $n=6$. 
\item
This shows that
the effective temperature of the KK gravitons is well below
the diffuse gamma-ray bound, even before any dilution
necessary to solve the radion moduli problem.  It also
demonstrates that the vast majority of the energy in the
bulk is in the motion of the zero mode radion $\rho_b$, rather
than in the bulk KK modes.  This is simply because
that $\rho_b\simeq \rho_{wall}$ is the natural circumstance
at the end of the epoch of contraction.
\end{itemize}

We now turn to the radion moduli problem:

\begin{itemize}
\item
Finally around the stabilization point $b_0$ the radion field
starts to oscillate freely.  Since this energy density scales as
$1/a^3$, and the wall-to-radion energy densities are initially
comparable at the start of the oscillation era, the radion
energy eventually dominates the total energy density.
\item
The most serious question that early universe cosmology presents
in the world-as-a-brane scenario is how do we dilute this energy
in radion oscillations to an acceptable level.  The radion is
long-lived, its' decay width back to light wall states being given
by\footnote{This is increased if there are many branes in the
bulk.  The ``brane crystallization"
mechanism of stabilization requires
$N_{\rm wall} \simeq (\mpl/\mst)^{2(n-2)/n}$ branes in the bulk.
If each of these have $O(1)$ light modes then the total decay
width to all branes is greatly enhanced.}
\be
\Ga_{\phi} \simeq {m_\phi^3\over \mpl^2}.
\ee
We thus require some dilution in the radion
energy density, either by a short period of late
inflation followed by reheating, or by a delayed reheating
after $\rho_{b}$ has sufficiently red-shifted. 
The amount of dilution of the radion energy density
that we require is relatively modest, given roughly by
$10^{-7}$, so that only about 5 efolds of
late inflation are needed.  
\end{itemize}

\section*{Conclusions}

We have argued that early inflation when the internal
dimensions are still small can successfully accomplish
all that is required of inflation, including generation
of suitable $\de\rho/\rho$ without the unpleasant introduction
of very light or fine-tuned wall fields.  Remarkably the era of
post-inflation brane-contraction is harmless, and automatically
ends via a ``Big Bounce''.
During the phase of $b(t)$ evolution to the
stabilization point, the production of bulk
gravitons by the time-varying metric remains completely suppressed,
ensuring that the bulk is very cold at, and after, the stabilization
of the internal dimensions.  The primary remaining issue is the radion
moduli problem, which is no more severe than in gauge-mediated
supersymmetry breaking models.  Overall, then, early universe cosmology
in these models is quite interesting!

\def\pl#1#2#3{{\it Phys. Lett. }{\bf B#1~}(19#2)~#3}
\def\zp#1#2#3{{\it Z. Phys. }{\bf C#1~}(19#2)~#3}
\def\prl#1#2#3{{\it Phys. Rev. Lett. }{\bf #1~}(19#2)~#3}
\def\rmp#1#2#3{{\it Rev. Mod. Phys. }{\bf #1~}(19#2)~#3}
\def\prep#1#2#3{{\it Phys. Rep. }{\bf #1~}(19#2)~#3}
\def\pr#1#2#3{{\it Phys. Rev. }{\bf D#1~}(19#2)~#3}
\def\np#1#2#3{{\it Nucl. Phys. }{\bf B#1~}(19#2)~#3}
\def\mpl#1#2#3{{\it Mod. Phys. Lett. }{\bf #1~}(19#2)~#3}
\def\arnps#1#2#3{{\it Annu. Rev. Nucl. Part. Sci. }{\bf #1~}(19#2)~#3}
\def\sjnp#1#2#3{{\it Sov. J. Nucl. Phys. }{\bf #1~}(19#2)~#3}
\def\jetp#1#2#3{{\it JETP Lett. }{\bf #1~}(19#2)~#3}
\def\app#1#2#3{{\it Acta Phys. Polon. }{\bf #1~}(19#2)~#3}
\def\rnc#1#2#3{{\it Riv. Nuovo Cim. }{\bf #1~}(19#2)~#3}
\def\ap#1#2#3{{\it Ann. Phys. }{\bf #1~}(19#2)~#3}
\def\ptp#1#2#3{{\it Prog. Theor. Phys. }{\bf #1~}(19#2)~#3}


\begin{thebibliography}{99}

\bibitem{ADD}
N. Arkani-Hamed, S. Dimopoulos and G. Dvali,
{\it Phys. Lett.} {\bf B429}, 263 (1998).

\bibitem{AADD}
I. Antoniadis, N. Arkani-Hamed, S. Dimopoulos and G. Dvali,
{\it Phys. Lett.} {\bf B436}, 257 (1998).

\bibitem{ADDlong}
N. Arkani-Hamed, S. Dimopoulos and G. Dvali, hep-ph/9807344.


\bibitem{antoniadis}
I. Antoniadis, \pl{246}{90}{377};
I. Antoniadis and K.~Benakli, \pl{326}{94}{69};
I. Antoniadis, K.~Benakli and M.~Quiros, \pl{331}{94}{313}.

\bibitem{other}
P. Horava and E. Witten, \np{460}{96}{506};
E. Witten \np{471}{96}{135};
J.D. Lykken, \pr{54}{96}{3693};
E. Caceres, V.S. Kaplunovsky and I.M.Mandelberg, \np{493}{97}{73}.

\bibitem{AHDMR}
N. Arkani-Hamed, S. Dimopoulos and J. March-Russell, hep-th/9809124.

\bibitem{new}
K. Dienes, E. Dudas and T. Gherghetta, {\it Phys. Lett.} {\bf B436},
55 (1998); \np{537}{99}{47};
G. Shiu and S.H. Tye, {\it Phys. Rev.} {\bf D58}, 106007 (1998);
C. Bachas, hep-ph/9807415;
R. Sundrum, hep-ph/9805471 and hep-ph/9807348;
P.~Argyres, S.~Dimopoulos and J.~March-Russell, hep-th/9808138;
Z. Kakushadze and S.-H. Tye,  hep-th/9809147;
K. Dienes, {\it et al.}, hep-ph/9809406;
K. Benakli, hep-ph/9809582;
L. Randall and R. Sundrum, hep-th/9810155;
G.F. Giudice, R. Rattazzi and J.D. Wells, hep-ph/9811291;
S. Nussinov and R. Shrock, hep-ph/9811323;
E.A. Mirabelli, M. Perelstein and M.E. Peskin, hep-ph/9811337;
T. Han, J.D. Lykken and R. Zhang, hep-ph/9811350;
N. Arkani-Hamed and S. Dimopoulos, hep-ph/9811353;
J.L. Hewett, hep-ph/9811356;
Z. Berezhiani and G. Dvali, hep-ph/9811378;
K.R. Dienes, E. Dudas and T. Gherghetta, hep-ph/9811428;
N. Arkani-Hamed, {\it et al.}, hep-ph/9811448;
I. Antoniadis and C. Bachas, hep-th/9812093;
Z. Kakushadze, hep-th/9812163.

\bibitem{cosmo}
K. Benakli and S. Davidson, hep-ph/9810280;
D. Lyth, hep-ph/9810320.

\bibitem{kl}
N. Kaloper and A. Linde, hep-th/9811141.

\bibitem{dt} 
G. Dvali and S.H. Tye, hep-ph/9812483.

\bibitem{Dbrane}
See for example: J. Polchinski, {\it TASI
lectures on D-branes}, hep-th/9611050;
C. Bachas, {\it Lectures on D-branes}, hep-th/9806199.

\bibitem{AHDKMR}
N. Arkani-Hamed, S. Dimopoulos, N. Kaloper, and
J. March-Russell, hep-ph/9803224.

\bibitem{nemanja}
N. Kaloper, talk at COSMO98, and contribution to this volume.

\end{thebibliography}
\end{document}